\documentclass[reqno]{amsart}

\newcommand{\Rm}{\mathbb{R}}

\newcommand{\Nm}{\mathbb{N}}

\newcommand{\be}{\[}
\newcommand{\ee}{\]}
\newcommand{\ba}{\[\begin{aligned}}
\newcommand{\ea}{\end{aligned}\]}

\newcommand{\pp}{\partial}
\newcommand{\bv}[1]{\boldsymbol{\mathrm{#1}}}

\usepackage{amsmath}
\usepackage[foot]{amsaddr}
\usepackage{amsthm,amssymb,mathrsfs}
\usepackage{graphicx}

\newtheorem{thm}{Theorem}[section]
\newtheorem{lem}[thm]{Lemma}

\theoremstyle{remark}\newtheorem{rmk}[thm]{Remark}

\title[]{Diffuse optical tomography in time domain with the inverse Rytov series}

\author[]{Chi Zhang$^1$ and Manabu Machida$^{2,*}$}
\address[]{$^1$Department of Cellular and Molecular Anatomy, 
Hamamatsu University School of Medicine, Hamamatsu 431-3192, Japan\\
$^2$Department of Informatics, Faculty of Engineering, 
Kindai University, Higashi-Hiroshima 739-2116, Japan}
\email{$^*$machida@hiro.kindai.ac.jp}

\begin{document}

\begin{abstract}
The Rytov approximation has been commonly used to obtain reconstructed images for optical tomography. However, the method requires linearization of the nonlinear inverse problem. Here, we demonstrate nonlinear Rytov approximations by developing the inverse Rytov series for the time-dependent diffusion equation. The method is verified by a solid-phantom experiment.
\end{abstract}

\maketitle

\section{Introduction}
\label{intro}

Optical tomography obtains reconstructed images similar to the X-ray computed tomography \cite{Boas-etal01}. However, the inverse problem for optical tomography becomes nonlinear and severely ill-posed because near-infrared light, which is used for optical tomography, is multiply scattered in biological tissue \cite{Arridge99}. One way to obtain reconstructed images of optical tomography is to solve the minimization problem for a cost function by an iterative scheme. Such iterative methods do not work especially for clinical research, in which less a priori knowledge is available compared with phantom experiments; choosing a good initial guess is difficult and the calculation is trapped by a local minimum since the cost function of optical tomography has a complicated landscape with local minima. The other way is to directly reconstruct perturbation of a coefficient. The Born and Rytov approximations are known in the direct approach. The Rytov approximation has been used in practical situations including optical tomography for the breast cancer \cite{Choe-etal05} and brain function \cite{Eggebrecht-etal14}. It was numerically demonstrated that the Rytov approximation appeared superior \cite{Arridge01}. The drawback of the (first) Born and Rytov approximations is that these methods require linearization of nonlinear inverse problems. That is, nonlinear terms in the Born and Rytov series are ignored.

In this paper, we will develop the latter approach of perturbation and consider nonlinear Rytov approximations. Although the Rytov approximation has been commonly used in optical tomography, it was only recently devised how to solve inverse problems by taking nonlinear terms in the Rytov series into account \cite{Machida23}.

The inverse Born series has been developed to invert the Born series \cite{Moskow-Schotland19}. The inverse Born series was considered for the Helmholtz equation \cite{Tsihrintzis-Devaney00a}, the diffusion equation \cite{Markel-OSullivan-Schotland03,Markel-Schotland07}, and the inverse scattering problem \cite{Panasyuk-Markel-Carney-Schotland06}. Its mathematical properties and recursive algorithm were developed \cite{Moskow-Schotland08,Moskow-Schotland09}. Furthermore, the inverse Born series was studied for the Calder\'{o}n problem \cite{Arridge-Moskow-Schotland12}, scalar waves \cite{Kilgore-Moskow-Schotland12}, the inverse transport problem \cite{Machida-Schotland15}, electromagnetic scattering \cite{Kilgore-Moskow-Schotland17}, discrete inverse problems \cite{Chung-Gilbert-Hoskins-Schotland17}, and the Bremmer series \cite{Shehadeh-Malcolm-Schotland17}. In \cite{Bardsley-Vasquez14,Lakhal18}, the inverse Born series was extended to Banach spaces. In \cite{Abhishek-Bonnet-Moskow20}, a modified Born series with unconditional convergence was proposed and its inverse series was studied. In \cite{Hoskins-Schotland22}, the convergence theorem for the inverse Born series has recently been improved. A reduced inverse Born series was proposed \cite{Markel-Schotland22}. The inverse Born series was extended to a nonlinear equation \cite{DeFilippis-etal23}. Its convergence, stability, and approximation error were proved under $H^s$ norm \cite{Mahankali-Yang23}.

The comparison of the Born and Rytov approximations has been discussed \cite{Keller69,Kirkinis08}. It is known that better reconstructed images can be obtained by the Rytov approximation. To extend the Rytov approximation, the inversion of the Rytov series has been studied. In \cite{Tsihrintzis-Devaney00b}, the inversion for the Helmholtz equation was performed but no general way of considering nonlinear terms was obtained. In \cite{Park-etal11}, the inversion of the Rytov series was studied but each term in the obtained series contains infinitely many higher-order terms and numerical reconstruction based on the obtained series was not feasible. In \cite{Machida23}, the inverse Rytov series was constructed to invert the Rytov series. Each term in the inverse Rytov series can be recursively computed. In this paper, by developing \cite{Machida23}, we will consider the inverse Rytov series for diffuse optical tomography in time domain and furthermore verify the inverse series experimentally.

The rest of the paper is organized as follows. In Sec.~\ref{fwd}, the diffuse light is expressed in the form of a series. In particular, the numerical algorithm for nonlinear Rytov approximations is explained in Sec.~\ref{ip}. In Sec~.\ref{phantomexp}, the experimental setup is described. To handle experimental data, we consider another series by taking difference of the inverse series. Results are shown in Sec.~\ref{results}. Section \ref{concl} is devoted to the concluding remarks.

\section{Forward and inverse series for diffuse light}
\label{fwd}

\subsection{Time-dependent diffusion equation}

Let $\Omega$ be the half-space in $\Rm^3$. The boundary of $\Omega$ is denoted by $\pp\Omega$. Let $c$ be the speed of light in $\Omega$. Let $x_s\in\Omega$ be the position of the source. The energy density $u(x,t;x_s)$ ($x\in\Omega$, $t\in(0,T)$) of near-infrared light in biological tissue is governed by the following diffusion equation \cite{Arridge99,Arridge-Schotland09}.

\be
\left\{\begin{aligned}
&\pp_tu-D_0\Delta u+\alpha u=S,\quad (x,t)\in\Omega\times(0,T),
\\
&u=0,\quad (x,t)\in\pp\Omega\times(0,T),
\\
&u=0,\quad x\in\Omega,\;t=0,
\end{aligned}\right.
\ee
where $S(x,t)$ is the source term and $D_0=c/(3\mu_s')>0$ is the diffusion coefficient with the reduced scattering coefficient $\mu_s'$. The positive constant $\alpha=c\mu_a$ is given by $c$ and the absorption coefficient $\mu_a$. Let $\delta(\cdot)$ be Dirac's delta function. The source term is given by
\be
S(x,t)=g\delta(x-x_s)h(t),
\ee
where $h(t)$ is the temporal profile of the light source, $g>0$ is a constant. We suppose that measurements are performed at $M_{\rm SD}$ places on $\pp\Omega$: $x_s=x_s^{(i)}=(x_{s1}^{(i)},x_{s2}^{(i)},\ell)$ for $i=1,\dots,M_{\rm SD}$. The positive constant $\ell$ will be set to $\ell=1/\mu_s'$. We write $\alpha$ as
\be
\alpha(x)=\left(1+\eta(x)\right)\alpha_0
\ee
with constant $\alpha_0>0$. We assume that $\left.\eta\right|_{x\in\pp\Omega}=0$.

When the measurement of diffuse optical tomography is performed, the reflectance $\Gamma^{(i)}(t)$ or the energy current in the normal direction is observed:
\be
\Gamma^{(i)}(t)=
-D_0\frac{\pp}{\pp x_3}u\left(x_d^{(i)},t;x_s^{(i)}\right),\quad 
x_d^{(i)}\in\pp\Omega.
\ee
Here we assumed that the out-going light at point $x_d^{(i)}\in\pp\Omega$ is observed for each source at $x_s^{(i)}$ ($i=1,\dots,M_{\rm SD}$).

Let $u_0$ be the solution to the following diffusion equation.
\begin{equation}
\left\{\begin{aligned}
&\pp_tu_0-D_0\Delta u_0+\alpha_0 u_0=S,\quad (x,t)\in\Omega\times(0,T),
\\
&u_0=0,\quad (x,t)\in\pp\Omega\times(0,T),
\\
&u_0=0,\quad x\in\Omega,\;t=0.
\end{aligned}\right.
\label{DE2}
\end{equation}
We introduce the Green's function $G(x,x';t-t')$ as the solution of (\ref{DE2}) when $S$ is replaced by $\delta(x-x')\delta(t-t')$. We obtain
\be
G(x,x';t)=
\frac{e^{-\alpha_0t}}{(4\pi D_0t)^{3/2}}
e^{-\frac{(x_1-x_1')^2+(x_2-x_2')^2}{4D_0t}}
\left(e^{-\frac{(x_3-x_3')^2}{4D_0t}}-e^{-\frac{(x_3+x_3')^2}{4D_0t}}\right),
\ee
for $t>0$ and $G=0$ if $t<0$. We note that
\be
\frac{\pp}{\pp x_3}G(x,x';t)=
\frac{e^{-\alpha_0t}}{(4\pi D_0)^{3/2}D_0}t^{-5/2}x_3'
e^{-\frac{(x_1-x_1')^2+(x_2-x_2')^2+(x_3')^2}{4D_0t}}
\ee
for $x\in\pp\Omega$, $x'\in\Omega$, $0<t<T$. Using the Green's function, $u_0$ can be written as
\be
u_0(x,t;x_s)=g\int_0^tG(x,x_s;t-s)h(s)\,ds.
\ee
We introduce operator $K_n$ ($n=1,2,\dots$):
\ba
\left(K_n\eta\otimes\dots\otimes\eta\right)(x,t;x_s)
&=
(-1)^{n+1}\alpha_0^n\int_0^T\int_{\Omega}\cdots\int_0^T\int_{\Omega}
G(x,y_1;t-s_1)\eta(y_1)
\\
&\times
G(y_1,y_2;s_1-s_2)\eta(y_2)\times\cdots\times G(y_{n-1},y_n;s_{n-1}-s_n)
\\
&\times
\eta(y_n)u_0(y_n,s_n;x_s)
\,dy_1ds_1\cdots dy_nds_n
\ea
for $n=1,2,\dots$. Here, the symbol $\otimes$ means tensor product \cite{Ryan02}.

Since the measured quantity $\Gamma^{(i)}$ contains the derivative of $u$, we define
\be
w(x_d,t;x_s)=\frac{\pp}{\pp x_3}u(x_d,t;x_s),\quad
x_d\in\pp\Omega,\quad x_s\in\Omega,\quad 0<t<T.
\ee
We write
\be
w_0(x_d,t;x_s)=g\int_0^t\frac{\pp G}{\pp x_3}(x_d,x_s;t-s)h(s)\,ds.
\ee
Correspondingly, we introduce
\ba
\left(K'_n\eta\otimes\dots\otimes\eta\right)(x_d,t;x_s)
&=
(-1)^{n+1}\alpha_0^n
\int_0^T\int_{\Omega}\cdots\int_0^T\int_{\Omega}
\frac{\pp G}{\pp x_3}(x_d,y_1;t-s_1)\eta(y_1)
\\
&\times
G(y_1,y_2;s_1-s_2)\eta(y_2)
\times\cdots\times
G(y_{n-1},y_n;s_{n-1}-s_n)
\\
&\times
\eta(y_n)u_0(y_n,s_n;x_s)\,dy_1ds_1\cdots dy_nds_n
\ea
for $n=1,2,\dots$.

The following lemma can be proved.

\begin{lem}
For sufficiently small $\eta$, $u$ can be expressed as
\be
u(x,t;x_s)=u_0(x,t;x_s)
-\sum_{n=1}^{\infty}(K_n\eta\otimes\dots\otimes\eta)(x,t;x_s),
\ee
and $w$ can be expressed as
\be
w(x_d,t;x_s)=w_0(x_d,t;x_s)
-\sum_{n=1}^{\infty}(K'_n\eta\otimes\dots\otimes\eta)(x_d,t;x_s).
\ee
\end{lem}

\begin{proof}
By subtraction, we have
\be
u(x,t;x_s)-u_0(x,t;x_s)
=
-\alpha_0\int_0^t\int_{\Omega}G(x,x';t-t')\eta(x')u(x',t';x_s)\,dx'dt'.
\ee
Thus we can construct the Born series as
\be
u=u_0+u_1+\cdots,
\ee
where
\be
u_n(x,t;x_s)=-\alpha_0
\int_0^t\int_{\Omega}G(x,x';t-t')\eta(x')u_{n-1}(x',t';x_s)\,dx'dt'
\ee
for $n=1,2,\dots$. We can write
\be
u_n(x,t;x_s)=-\left(K_n\eta\otimes\dots\otimes\eta\right)(x,t;x_s)
\ee
for $n=1,2,\dots$. Thus the first part of the lemma is proved. The last half of the lemma can be proved as follows.

We consider another Born series as
\be
w=w_0+w_1+\cdots.
\ee
We obtain for $n\ge1$,
\be
w_n(x_d,t;x_s)=
-\alpha_0\int_0^t\int_{\Omega}\frac{\pp G}{\pp x_3}(x_d,x';t-t')\eta(x')
u_{n-1}(x',t';x_s)\,dx'dt'.
\ee
We express $w_n$ ($n=1,2,\dots$) as
\be
w_n(x_d,t;x_s)=-\left(K'_n\eta\otimes\dots\otimes\eta\right)(x_d,t;x_s).
\ee
\end{proof}

\begin{rmk}
See \cite{Moskow-Schotland08} for the convergence in Lemma 2.1.
\end{rmk}

\begin{rmk}
We note the relation
\be
\left(K'_n\eta\otimes\dots\otimes\eta\right)(x_d,t;x_s)
=
-\alpha_0\int_0^t\int_{\Omega}\frac{\pp G}{\pp x_3}(x_d,y;t-s)
\eta(y)
\left(K_{n-1}\eta\otimes\dots\otimes\eta\right)(y,s;x_s)\,dyds.
\ee
\end{rmk}

The first-order term in the Born series is obtained as the first-order perturbation. This is called the (first) Born approximation. By applying the Born approximation, we can solve nonlinear inverse problems with the pseudoinverse. It is known that the Rytov approximation, which is another first-order approximation, gives better reconstructed images than the Born approximation \cite{Arridge01}. This motivates us to investigate the Rytov and inverse Rytov series.

Let us write
\be
\psi=\ln\frac{w_0}{w}.
\ee
We define
\ba
&
\left(J_n\eta\otimes\dots\otimes\eta\right)(x_d,t;x_s)=
\sum_{m=1}^n\frac{1}{mw_0(x_d,t;x_s)^n}
\\
&\times
\sum_{j_1+\cdots+j_m=n}\left((K'_{j_1}\eta\otimes\dots\otimes\eta)(x_d,t;x_s)\right)
\times\cdots\times
\left((K'_{j_m}\eta\otimes\dots\otimes\eta)(x_d,t;x_s)\right).
\ea

\begin{thm}[Rytov series]
Assume that $\eta$ is sufficiently small. Then $\psi$ can be expressed as
\be
\psi(x_d,t;x_s)=\sum_{n=1}^{\infty}(J_n\eta\otimes\dots\otimes\eta)(x_d,t;x_s).
\ee
\end{thm}

\begin{proof}
Let us write $w$ as
\be
w=w_0e^{-\psi_1-\psi_2-\cdots}.
\ee
By expanding $w$ in $-\ln(w/w_0)$ with the Born series, we have
\be
\psi_n(x_d,x_s;t)
=
\sum_{m=1}^n\frac{(-1)^m}{mw_0(x_d,t;x_s)^m}
\sum_{j_1+\cdots+j_m=n}w_{j_1}(x_d,t;x_s)\cdots w_{j_m}(x_d,t;x_s)
\ee
for $n=1,2,\dots$. Hence we can write
\be
\psi_n(x_d,x_s;t)=\left(J_n\eta\otimes\dots\otimes\eta\right)(x_d,t;x_s).
\ee
\end{proof}

\subsection{Recursive algorithm for the inverse series}
\label{ip}

Considering the experiment in Sec.~\ref{phantomexp}, we assume that $\eta(x)$ is independent of $x_2$ and write $\eta=\eta(x_1,x_3)$. The light emitted at $x_s=(x_{s1},0,\ell)$ is detected at $x_d=(x_{d1},0,0)$. For spatial variables $y_1,y_3$ and time $t$, discretization is done as follows:
\be
y_1^{(l_1)}=l_1\Delta x,\quad
y_3^{(l_3)}=l_3\Delta x,\quad
t^{(k)}=(k-1)\Delta t,
\ee
where $\Delta t$ ($>0$) is the temporal resolution of measurements, $\Delta x>0$, and
\be
1\le l_1\le N_1,\quad 1\le l_3\le N_3,\quad 1\le k\le N_T.
\ee
In our optical tomography, $(N_1,N_3,\Delta x)=(60,40,0.5\,{\rm mm})$, $(30,20,1\,{\rm mm})$, or $(15,10,2\,{\rm mm})$. Moreover, $\Delta t=10\,{\rm ps}$, $N_T=200$.

Let us define vectors $\bv{K}_0\in\Rm^{M_{\rm SD}N_T}$, $\bv{K}'_0\in\Rm^{M_{\rm SD}N_T}$, $\bv{K}'_1\in\Rm^{M_{\rm SD}N_T}$, $\bv{K}_1\in\Rm^{N_1N_3M_{\rm SD}N_T}$ as
\ba
\{\bv{K}_0\}_{k+(i-1)N_T}
&=
-g\int_0^{t^{(k)}}G\left(x_{d1}^{(i)},0,0^+,x_{s1}^{(i)},0,0^+;t^{(k)}-s\right)h(s)\,ds,
\\
\{\bv{K}'_0\}_{k+(i-1)N_T}
&=
-g\int_0^{t^{(k)}}\frac{\pp G}{\pp x_3}\left(x_{d1}^{(i)},0,x_{s1}^{(i)},x_{s2}^{(i)},\ell;t^{(k)}-s\right)h(s)\,ds,
\ea
\ba
\{\bv{K}'_1(\bv{b})\}_{k+(i-1)N_T}
&=
\alpha_0(\Delta x)^2
\sum_{l_1'=1}^{N_1}\sum_{l_3'=1}^{N_3}
\int_0^T\frac{\pp G}{\pp x_3}
\left(x_{d1}^{(i)},0,0,y_1^{(l_1')},0,y_3^{(l_3')};t^{(k)}-s\right)
\\
&\times
\{\bv{b}\}_{l_1'+(l_3'-1)N_1}
u_0\left(y_1^{(l_1')},0,y_3^{(l_3')},x_{s1}^{(i)},0,\ell;s\right)\,ds,
\ea
and
\ba
&
\{\bv{K}_1(\bv{b})\}_{k+(i-1)N_T+(l_1-1)M_{\rm SD}N_T+(l_3-1)N_1M_{\rm SD}N_T}
\\
&=
\alpha_0(\Delta x)^2\sum_{l_1'=1}^{N_1}\sum_{l_3'=1}^{N_3}
\int_0^T
G\left(x_1^{(l_1)},0,x_3^{(l_3)},y_1^{(l_1')},0,y_3^{(l_3')};t^{(k)}-s\right)
\\
&\times
\{\bv{b}\}_{l_1'+(l_3'-1)N_1}
u_0\left(y_1^{(l'_1)},0,y_3^{(l'_3)},x_{s1}^{(i)},0,\ell;s\right)\,ds.
\ea
For $n\ge2$, we define $\bv{K}'_n(\bv{b}_1,\dots,\bv{b}_n)\in\Rm^{M_{\rm SD}N_T}$, $\bv{K}_n(\bv{b}_1,\dots,\bv{b}_n)\in\Rm^{N_1N_3M_{\rm SD}N_T}$ as
\ba
\{\bv{K}'_n(\bv{b}_1,\dots,\bv{b}_n)\}_{k+(i-1)N_T}
&=
-\alpha_0(\Delta x)^2\Delta t
\sum_{k'=1}^{N_T}\sum_{l_1'=1}^{N_1}\sum_{l_3'=1}^{N_3}\frac{\pp G}{\pp x_3}
\left(x_{d1}^{(i)},0,0,y_1^{(l_1')},0,y_3^{(l_3')};
t^{(k)}-t^{(k')}\right)
\\
&\times
\{\bv{b}_n\}_{l_1'+(l_3'-1)N_1}\{\bv{K}_{n-1}(\bv{b}_1,\dots,\bv{b}_{n-1})\}_{\widetilde{m}},
\ea
and
\ba
&
\{\bv{K}_n(\bv{b}_1,\dots,\bv{b}_n)\}_{k+(i-1)N_T+(l_1-1)M_{\rm SD}N_T+(l_3-1)N_1M_{\rm SD}N_T}
\\
&=
-\alpha_0(\Delta x)^2\Delta t
\sum_{k'=1}^{N_T}\sum_{l_1'=1}^{N_1}\sum_{l_3'=1}^{N_3}
G\left(x_1^{(l_1)},0,x_3^{(l_3)},y_1^{(l_1')},0,y_3^{(l_3')};t^{(k)}-t^{(k')}\right)
\\
&\times
\{\bv{b}_n\}_{l_1'+(l_3'-1)N_1}
\{\bv{K}_{n-1}(\bv{b}_1,\dots,\bv{b}_{n-1})\}_{\widetilde{m}}.
\ea
Here, we introduced the notation
\be
\widetilde{m}=k'+(i-1)N_T+(l_1'-1)M_{\rm SD}N_T+(l_3'-1)N_1M_{\rm SD}N_T.
\ee
We set
\be
n_t=n_2-n_1+1,
\ee
where $t_1=n_1\Delta t$ and $t_2=n_2\Delta t$. We introduce $\bv{J}_n(\bv{b}_1,\dots,\bv{b}_n)\in\Rm^{M_{\rm SD}(n_t+1)}$ as
\ba
&
\{\bv{J}_n(\bv{b}_1,\dots,\bv{b}_n)\}_{k-n_1+1+(i-1)(n_t+1)}
=
\sum_{m=1}^n\frac{(-1)^m}{m\{\bv{K}'_0\}_{k+(i-1)N_T}^m}
\\
&\times
\sum_{j_1+\cdots+j_m=n}
\{\bv{K}'_{j_1}(\bv{b}_1,\dots,\bv{b}_{j_1})\}_{k+(i-1)N_T}
\times\cdots\times
\{\bv{K}'_{j_m}(\bv{b}_{n-j_m+1},\dots,\bv{b}_n)\}_{k+(i-1)N_T}
\ea
for $k=n_1,\dots,n_2+1$. We note that
\be
\{\bv{J}_1(\bv{b})\}_{k-n_1+1+(i-1)(n_t+1)}=
-\frac{1}{\{\bv{K}'_0\}_{k+(i-1)N_T}}
\{\bv{K}'_1(\bv{b})\}_{k+(i-1)N_T}
\ee
for $k=n_1,\dots,n_2+1$. Let us define matrix $\underline{J}_1\in\Rm^{M_{\rm SD}(n_t+1)\times N_1N_3}$ as
\ba
&
\{\underline{J}_1\}_{k-n_1+1+(i-1)(n_t+1),l_1+(l_3-1)N_1}
=
-\frac{\alpha_0(\Delta x)^2}{\{\bv{K}'_0\}_{k+(i-1)N_T}}
\\
&\times
\int_0^T\frac{\pp G}{\pp x_3}
\left(x_{d1}^{(i)},0,0,y_1^{(l_1)},0,y_3^{(l_3)};t^{(k)}-s\right)
u_0\left(y_1^{(l_1)},0,y_3^{(l_3)},x_{s1}^{(i)},x_{s2}^{(i)},\ell;s\right)\,ds
\ea
for $k=n_1,\dots,n_2+1$, $i=1,\dots,M_{\rm SD}$. Then we can write
\be
\bv{J}_1(\bv{\eta})=\underline{J}_1\bv{\eta}.
\ee
Let $\underline{\mathcal{J}}_1\in\Rm^{N_1N_3\times M_{\rm SD}(n_t+1)}$ be a regularized pseudoinverse of $\underline{J}_1$.

Let us define vector $\bv{\Psi}$ as
\be
\{\bv{\Psi}\}_{k-n_1+1+(i-1)(n_t+1)}=
\ln\frac{C_{0i}w_0\left(x_1^{(i)},0,0,t^{(k)}\right)}{C_iw\left(x_1^{(i)},0,0,t^{(k)}\right)}
\ee
for $k=n_1,\dots,n_2+1$, $i=1,\dots,M_{\rm SD}$. By taking the time difference, we obtain
\be
\{\bv{\phi}\}_{k-n_1+1+(i-1)n_t}=
\{\bv{\Psi}\}_{k-n_1+2+(i-1)(n_t+1)}-
\{\bv{\Psi}\}_{k-n_1+1+(i-1)(n_t+1)}
\ee
for $k=n_1,\dots,n_2$, $i=1,\dots,M_{\rm SD}$. We have ($k=1,\dots,n_t$, $i=1,\dots,M_{\rm SD}$)
\be
\{\bv{\phi}\}_{k+(i-1)n_t}=
\{\bv{I}_1(\bv{\eta})\}_{k+(i-1)n_t}+\{\bv{I}_2(\bv{\eta})\}_{k+(i-1)n_t}
+\cdots.
\ee
Here, for each $n=1,2,\dots$,
\be
\{\bv{I}_n(\bv{\eta})\}_{k-n_1+1+(i-1)n_t}=
\{\bv{J}_n(\bv{\eta})\}_{k-n_1+2+(i-1)(n_t+1)}-
\{\bv{J}_n(\bv{\eta})\}_{k-n_1+1+(i-1)(n_t+1)}
\ee
for $k=n_1,\dots,n_2$, $i=1,\dots,M_{\rm SD}$. The matrix $\underline{I}_1\in\Rm^{M_{\rm SD}n_t\times N_1N_3}$ is introduced similarly:
\ba
\{\underline{I}_1\}_{k-n_1+1+(i-1)n_t,l_1+(l_3-1)N_1}
&=
\{\underline{J}_1\}_{k-n_1+2+(i-1)(n_t+1),l_1+(l_3-1)N_1}
\\
&-
\{\underline{J}_1\}_{k-n_1+1+(i-1)(n_t+1),l_1+(l_3-1)N_1}
\ea
for $k=n_1,\dots,n_2$, $i=1,\dots,M_{\rm SD}$. Then we can write
\be
\bv{I}_1(\bv{\eta})=\underline{I}_1\bv{\eta}.
\ee
Let $\underline{\mathcal{I}}_1\in\Rm^{N_1N_3\times M_{\rm SD}n_t}$ be a regularized pseudoinverse of $\underline{I}_1$. We write
\be
\bv{\eta}_1=\underline{\mathcal{I}}_1\bv{\phi},
\ee
where
\be
\{\bv{\eta}_1\}_{l_1+(l_3-1)N_1}=\eta_1(x_1^{(l_1)},0,x_3^{(l_3)}).
\ee

To solve the inverse problem, we introduce
\be
\bv{\eta}_j^{(1)}=\underline{\mathcal{I}}_1\bv{a}_j\quad(j=1,\dots,n)
\ee
for vectors $\bv{a}_1,\dots,\bv{a}_n\in\Rm^{M_{\rm SD}n_t}$. The inverse series for the forward series $I_1+I_2+\cdots$ can be recursively constructed as follows. Let us introduce vectors $\bv{\mathcal{I}}(\bv{a}_1,\dots,\bv{a}_n)\in\Rm^{N_1N_3}$ which have a recursive structure:
\be
\bv{\mathcal{I}}_1(\bv{a}_1)=\bv{\eta}_1^{(1)},
\ee
\be
\bv{\mathcal{I}}_n(\bv{a}_1,\dots,\bv{a}_n)=
-\sum_{m=1}^{n-1}\sum_{j_1+\cdots+j_m=n}
\bv{\mathcal{I}}_m\left(
\bv{I}_{j_1}(\bv{\eta}_1^{(1)},\dots,\bv{\eta}_{j_1}^{(1)}),\dots,
\bv{I}_{j_m}(\bv{\eta}_{n-j_m+1}^{(1)},\dots,\bv{\eta}_n^{(1)})\right)
\ee
for $n\ge2$. We note that the number of compositions for $j_1+\cdots+j_m=n$ is $(n-1)!/[(m-1)!(n-m)!]$.

Let us express the $N$th-order Rytov approximation as
\begin{equation}
\bv{\eta}^{(N)}=\bv{\eta}_1+\cdots+\bv{\eta}_N,
\label{reconN}
\end{equation}
where
\be
\bv{\eta}_n=\bv{\mathcal{I}}_n(\bv{\psi},\dots,\bv{\psi}),\quad
n=1,\dots,N.
\ee
Then the reconstructed absorption coefficient is written as
\be
\mu_a(x_1^{(l_1)},0,x_3^{(l_3)})=
\left(1+\{\bv{\eta}^{(N)}\}_{l_1+(l_3-1)N_1}\right)\mu_{a0},
\ee
where $\mu_{a0}=\alpha_0/c$.

\subsection{Time-independent diffusion equation}
\label{tindep}

The reconstruction (\ref{reconN}) requires fewer $N$ if the perturbation $\eta$ is smaller. In \cite{Machida23}, the convergence of the inverse series was numerically investigated for different values of $\eta$. Indeed, the convergence also depends on $|\omega|$, i.e., the size of the support of $\eta$. Here we study this with a relatively simple radial problem.

Let $\Omega\subset\Rm^2$ a disk of radius $R$ centered at the origin. In the polar coordinate system we have $x=(r,\theta)$. The diffusion equation is given by
\be
\left\{\begin{aligned}
&-D_0\Delta u+\alpha u=S,\quad x\in\Omega,
\\
&\ell\pp_{\nu}u+u=0,\quad x\in\pp\Omega,
\end{aligned}\right.
\ee
where the Robin boundary condition is imposed with a constant $\ell>0$ and the directional derivative $\pp_{\nu}$ with the outward unit vector $\nu$ normal to $\pp\Omega$. The source term $S$ is given by
\be
S(r,\theta)=e^{ik\theta}\frac{1}{r}\delta(r-R),
\ee
where $k\in\Nm$. Assuming an absorber disk of radius $r_a$ ($<R$) in $\Omega$, we write $\alpha$ as
\be
\alpha(x)=\left(1+\eta(r)\right)\alpha_0
\ee
with constant $\alpha_0>0$. Here,
\be
\eta(r)=\left\{\begin{aligned}
\eta_a,&\quad0\le r\le r_a,
\\
0,&\quad r_a<r\le R.
\end{aligned}\right.
\ee
Suppose that we measure the detected light $u$ at $(r,\theta)=(R,0)$.

Let $u_0$ be the solution to the following diffusion equation.
\begin{equation}
\left\{\begin{aligned}
&-D_0\Delta u_0+\alpha_0u_0=
e^{ik\theta}\frac{1}{r}\delta(r-R),\quad x\in\Omega,
\\
&\ell\pp_{\nu}u_0+u_0=0,\quad x\in\pp\Omega.
\end{aligned}\right.
\label{DE2b}
\end{equation}
We introduce the Green's function $G(x,x')$ as the solution of (\ref{DE2b}) when the source term is replaced by $\frac{1}{r}\delta(r-r')\delta(\theta-\theta')$. Let us express the Green's function as
\be
G(x,x')=\frac{1}{2\pi}\sum_{n=-\infty}^{\infty}e^{in(\theta-\theta')}g_n(r,r').
\ee
Let us write
\be
\beta=\sqrt{\frac{\alpha_0}{D_0}}.
\ee
Noting that $\Delta=\pp_r^2+r^{-1}\pp_r+r^{-2}\pp_{\theta}^2$ and $\sum_{n=-\infty}^{\infty}e^{in\theta}=2\pi\delta(\theta)$, we find
\ba
r^2\pp_r^2g_n(r,r')+r\pp_rg_n(r,r')-\left(\beta^2r^2+n^2\right)g_n(r,r')
&=
-\frac{r}{D_0}\delta(r-r'),
\\
\ell\pp_rg_n(R,r')+g_n(R,r')&=0.
\ea

Let us change the variable as
\be
z=\beta r.
\ee
Then we have
\ba
z^2\pp_z^2g_n(r,r')+z\pp_zg_n(r,r')-\left(z^2+n^2\right)g_n(r,r')
&=
-\frac{z}{D_0}\delta(z-z'),
\quad 0<z<\beta R,
\\
\beta\ell\pp_zg_n(R,r')+g_n(R,r')&=0,
\ea
where $z'=\beta r'$. In this numerical test, we set
\be
D_0=1.
\ee
We can write
\be
g_n(r,r')=\left\{\begin{aligned}
AI_n(z),&\quad r<r',
\\
BI_n(z)+CK_n(z),&\quad r>r'.
\end{aligned}\right.
\ee
We have
\be
AI_n(z')=BI_n(z')+CK_n(z').
\ee
Moreover, the jump condition and boundary condition read
\ba
(B-A)\frac{dI_n}{dx}(z')+C\frac{dK_n}{dx}(z')=-\frac{1}{z'},
\\
\beta\ell\left(B\frac{dI_n}{dx}(\beta R)+C\frac{dK_n}{dx}(\beta R)\right)+BI_n(\beta R)+CK_n(\beta R)=0.
\ea
Coefficients are calculated as ($-K_nI_n'+I_nK_n'=-1/z'$)
\ba
A&=
K_n(z')-\frac{K_n(\beta R)+\beta\ell(dK_n/dx)(\beta R)}{I_n(\beta R)+\beta\ell(dI_n/dx)(\beta R)}I_n(z'),
\\
B&=
-\frac{K_n(\beta R)+\beta\ell(dK_n/dx)(\beta R)}{I_n(\beta R)+\beta\ell(dI_n/dx)(\beta R)}I_n(z'),
\\
C&=
I_n(z').
\ea
Thus,
\be
g_n(r,r')=K_n(\max(z,z'))I_n(\min(z,z'))
-\frac{K_n(\beta R)+\beta\ell(dK_n/dx)(\beta R)}{I_n(\beta R)+\beta\ell(dI_n/dx)(\beta R)}I_n(z)I_n(z').
\ee
We obtain
\be
g_n(r,r')=K_n(\beta\max(r,r'))I_n(\beta\min(r,r'))-d_n\frac{I_n(\beta r)I_n(\beta r')}{I_n(\beta R)},
\ee
where $I_n,K_n$ are modified Bessel functions of the first and second kinds, and
\be
d_n=\frac{K_n(\beta R)+\beta\ell K_n'(\beta R)}{I_n(\beta R)+\beta\ell I_n'(\beta R)}I_n(\beta R).
\ee
We obtain
\be
u_0(r,\theta)=u_0(r,\theta;k)=g_k(r,R)e^{ik\theta}.
\ee

Let us set
\be
\beta_a=\sqrt{\frac{(1+\eta_a)\alpha_0}{D_0}}.
\ee
The Green's function $G_a$ for $u$ can be written as
\be
G_a(z,z')=
\frac{1}{2\pi}\sum_{n=-\infty}^{\infty}a_ne^{in(\theta-\theta')}I_n(\beta_ar),
\quad r\in[0,r_a],
\ee
and
\ba
G_a(z,z')&=
\frac{1}{2\pi}\sum_{n=-\infty}^{\infty}e^{in(\theta-\theta')}I_n(\beta r)K_n(\beta R)
\\
&+
\frac{1}{2\pi}\sum_{n=-\infty}^{\infty}e^{in(\theta-\theta')}\left(
b_nK_n(\beta r)+c_nI_n(\beta r)\right),\quad r\in[r_a,R],
\ea
where the first term on the right-hand side is the fundamental solution with the source at $(R,\theta')$. Here, coefficients $a_n,b_n,c_n$ are computed as the solution to the following system of linear equations, which is derived from the interface and boundary conditions:
\ba
&
\begin{pmatrix}
I_n(\beta_a r_a) & -K_n(\beta r_a) & -I_n(\beta r_a) \\
\beta_aI_n'(\beta_ar_a) & -\beta K_n'(\beta r_a) & -\beta I_n'(\beta r_a) \\
0 & K_n(\beta R)+\beta\ell K_n'(\beta R) & I_n(\beta R)+\beta\ell I_n'(\beta R)
\end{pmatrix}
\begin{pmatrix}
a_n \\ b_n \\ c_n
\end{pmatrix}
\\
&=
\begin{pmatrix}
I_n(\beta r_a)K_n(\beta R) \\
\beta I_n'(\beta r_a)K_n(\beta R) \\
\beta\ell I_n(\beta R)K_n'(\beta R)+I_n(\beta R)K_n(\beta R)
\end{pmatrix}.
\ea
We obtain
\be
u(r,\theta)=u(r,\theta;k)=e^{ik\theta}
\left(I_k(\beta r)K_k(\beta R)+b_kK_k(\beta r)+c_kI_k(\beta r)\right),
\quad r_a<r<R.
\ee

Let $M$ be the number of spatial frequencies ($k=1,\dots,M$). We have
\ba
\psi_k
&=
\ln\frac{u_0(R,0;k)}{u(R,0;k)}
\\
&=
\ln\frac{\left(K_k(\beta R)-d_n\right)I_k(\beta R)}{I_k(\beta R)K_k(\beta R)+b_kK_k(\beta R)+c_kI_k(\beta R)},
\quad k=1,\dots,M.
\ea

The reconstruction can be done as follows. Let us introduce
\be
G^{(n)}(r,r')=g_n(r,r')r'.
\ee
We obtain
\ba
\left(K_n\eta^{\otimes n}\right)(x)
&=
(-1)^{n+1}\beta^{2n}e^{ik\theta}\int_0^R\cdots\int_0^R
G^{(k)}(R,r_1)G^{(k)}(r_1,r_2)
\\
&\times\cdots\times G^{(k)}(r_{n-1},r_n)G^{(k)}(r_n,R)
\eta(r_1)\cdots\eta(r_n)\,dr_1\dots d_n,\quad x\in\pp\Omega,
\ea
and define
\be
\left(\check{K}_n\eta^{\otimes n}\right)(x)=
\left(\frac{1}{u_0}K_n\eta^{\otimes n}\right)(x)=
\frac{1}{G^{(k)}(R,R)}\left(K_n\eta^{\otimes n}\right)(x),
\quad x\in\pp\Omega.
\ee
The Rytov series is expressed as
\be
J_n\eta^{\otimes n}=\sum_{m=1}^n\frac{1}{m}\sum_{j_1+\cdots+j_m=n}
\left(\check{K}_{j_1}\eta^{\otimes j_1}\right)\cdots
\left(\check{K}_{j_m}\eta^{\otimes j_m}\right).
\ee

We set
\be
r_i=i\Delta r\quad(i=1,\dots,N_r),\quad
\Delta r=\frac{R}{N_r}.
\ee
Let $\bv{b}\in\Rm^{N_r}$ be a vector. We define $\bv{K}_0\in\Rm^M$, $\bv{K}_1\in\Rm^{MN_r}$ as
\ba
\left\{\bv{K}_0\right\}_k&=-G^{(k)}(R,R),
\\
\left\{\bv{K}_1(\bv{b})\right\}_{i+(k-1)N_r}&=
\beta^2\Delta r\sum_{l=1}^{N_r}G^{(k)}(r_i,r_l)G^{(k)}(r_l,R)\{\bv{b}\}_l
\ea
for $1\le k\le M$, $1\le i\le N_r$. Moreover,
\ba
&\left\{\bv{K}_n(\bv{b}_1,\dots,\bv{b}_n)\right\}_{i+(k-1)N_r}
\\
&=
-\beta^2\Delta r\sum_{l=1}^{N_r}G^{(k)}(r_i,r_l)\{\bv{b}_n\}_l
\left\{\bv{K}_{n-1}(\bv{b}_1,\dots,\bv{b}_{n-1})\right\}_{l+(k-1)N_r}.
\ea
We introduce
\ba
\left\{\bv{J}_n(\bv{b}_1,\dots,\bv{b}_n)\right\}_k
&=
\sum_{m=1}^n\frac{(-1)^m}{m\{\bv{K}_0\}_k^m}
\\
&\times
\sum_{j_1+\cdots+j_m=n}\left\{\bv{K}_{j_1}(\bv{b}_1,\dots,\bv{b}_{j_1})\right\}_{kN_r}\cdots
\left\{\bv{K}_{j_m}(\bv{b}_{n-j_m+1},\dots,\bv{b}_n)\right\}_{kN_r}
\ea
for $k=1,\dots,M$.

Let us focus on the linear term:
\be
\left\{\bv{J}_1(\bv{b})\right\}_k=
-\frac{1}{\{\bv{K}_0\}_k}\left\{\bv{K}_1(\bv{b})\right\}_{kN_r}
\ee
for $k=1,\dots,M$, $i=1,\dots,N_r$. We can define a matrix $\underline{J}_1\in\Rm^{M\times N_r}$ such that $\bv{J}_1(\bv{b})=\underline{J}_1\bv{b}$ with elements
\be
\left\{\underline{J}_1\right\}_{k,i}=
\frac{\beta^2\Delta r}{G^{(k)}(R,R)}G^{(k)}(R,r_i)^2.
\ee
The linear inverse problem can be solved with $\underline{\mathcal{J}}_1$, which is a regularized Moore-Penrose pseudoinverse:
\be
\underline{\mathcal{J}}_1=\underline{J}_{1,{\rm reg}}^+\in\Rm^{N_r\times M}.
\ee
The observed data is stored in vector $\bv{\psi}=(\psi_k)\in\Rm^M$. The linear inverse problem is solved as
\be
\bv{\eta}_1=\underline{\mathcal{J}}_1\bv{\psi},
\ee
where $\bv{\eta}_1=(\eta_1(r_i))$ ($i=1,\dots,N_r$) is the solution within the linear (conventional) approximation. The calculation can be done as follows using the singular value decomposition. Below, we assume the underdetermined case ($M<N_r$) but the overdetermined case ($M>N_r$) can be considered similarly. We have
\be
\bv{\eta}_1=\underline{J}_1^*\left(\underline{J}_1\underline{J}_1^*\right)_{\rm reg}^{-1}\bv{\psi},
\ee
where $*$ denotes the Hermitian conjugate and $(\cdot)_{\rm reg}^{-1}$ means the pseudoinverse with regularization. We perform the regularization by discarding singular values that are smaller than a certain value $\sigma_0$. Let $\sigma_n^2$ and $\bv{v}_n$ be the eigenvalues and eigenvectors of the matrix $\underline{J}_1\underline{J}_1^*$:
\be
\left(\underline{J}_1\underline{J}_1^*\right)\bv{z}_n=\sigma_n^2\bv{z}_n.
\ee
We obtain
\be
\bv{\eta}_1=\sum_{n\atop\sigma_n\ge\sigma_0}\frac{1}{\sigma_n^2}
\left(\bv{z}_n^*\bv{\psi}\right)\underline{J}_1^*\bv{z}_n.
\ee

Let $\bv{a}_1,\dots\bv{a}_n$ be vectors in $\Rm^M$. Let us introduce
\be
\bv{\eta}_j^{(1)}=\underline{\mathcal{J}}_1\bv{a}_j\quad(j=1,\dots,n).
\ee
Moreover we introduce vector $\bv{\mathcal{J}}_n(\bv{a}_1,\dots,\bv{a}_n)\in\Rm^{N_r}$ which has a recursive structure:
\ba
\bv{\mathcal{J}}_1(\bv{a}_1)
&=
\bv{\eta}_1^{(1)},
\\
\bv{\mathcal{J}}_n(\bv{a}_1,\dots,\bv{a}_n)
&=
-\sum_{m=1}^{n-1}\sum_{j_1+\cdots+j_m=n}
\bv{\mathcal{J}}_m\left(\bv{J}_{j_1}(\bv{\eta}_1^{(1)},\dots,\bv{\eta}_{j_1}^{(1)}),\dots,\bv{J}_{j_m}(\bv{\eta}_{n-j_m+1}^{(1)},\dots,\bv{\eta}_n^{(1)})\right)
\ea
for $n\ge2$. We note that for each pair of $(n,m)$, there are $\begin{pmatrix}n-1\\ m-1\end{pmatrix}$ compositions which satisfy $j_1+\cdots+j_m=n$. Using the functions $\bv{\mathcal{J}}_n$, the $n$th term in the inverse Rytov series is can be computed as
\be
\bv{\eta}_n=\bv{\mathcal{J}}_n(\bv{\psi},\dots,\bv{\psi}).
\ee

Then the $N$th-order approximation is given by
\be
\bv{\eta}^{(N)}=\bv{\eta}_1+\cdots+\bv{\eta}_N.
\ee
The approximate function $\eta(r_i)$ can be drawn by plotting elements of the vector $\bv{\eta}^{(N)}$.

\subsection{Convergence of the inverse Rytov series}

To consider the convergence of the inverse Rytov series, we consider the following diffusion equation in a bounded domain $\Omega\subset\Rm^n$ ($n\ge2$) with a smooth boundary $\pp\Omega$.
\begin{equation}
\left\{\begin{aligned}
&\pp_tu-D_0\Delta u+(1+\eta)\alpha_0u=f,\quad x\in\Omega,\quad 0<t<T,
\\
&\ell\pp_{\nu}u+u=0,\quad x\in\pp\Omega,\quad 0<t<T,
\\
&u=0,\quad x\in\Omega,\quad t=0,
\end{aligned}\right.
\label{DE3}
\end{equation}
where $T$ is the observation time and $f=f(x,t)$ is the source term. We suppose that the support of $\eta$ is contained in a closed ball $\omega\subset\Omega$ and $\eta\in L^q(\omega)$ ($q\ge2$). For the inverse problems, measured data $u\in L^p(\pp\Omega)$ on the boundary is used ($p\ge1$). Let us introduce
\ba
\mu&=\alpha_0\sup_{x\in\omega}\|G(x,\cdot;\cdot)\|_{L^1(0,T;L^{q/(q-1)}(\omega))},
\\
\nu&=
\alpha_0T|\omega|^{1-1/q}\sup_{\substack{y_1,y_2\in\omega\\0<t<T\\0<s<T}}\left\|
G(\cdot,y_1;s)\frac{u_0(y_2,t)}{u_0(\cdot,t)}\right\|_{L^p(\pp\Omega)}.
\ea
Here, $u_0(x,t)$ is the solution of the diffusion equation (\ref{DE3}) in which $\eta$ is set to zero and $G(x,y;t)$ is the Green's function which satisfies (\ref{DE3}) when $\eta\equiv0$ and $f$ is replaced by $\delta(x-y)\delta(t)$. We note that the Born series can be written as
\be
u(x,t)=u_0(x,t)-\sum_{n=1}^{\infty}(K_n\eta\otimes\cdots\otimes\eta)(x,t),
\ee
where
\ba
&
(K_nf_1\otimes\cdots\otimes f_n)(x,t)
=(-1)^{n+1}\alpha_0^n\int_0^T\int_{\Omega}\cdots\int_0^T\int_{\Omega}
G(x,y_1;t-s_1)f_1(y_1)G(y_1,y_2;s_1-s_2)
\\
&\times
f_2(y_2)\times\cdots\times G(y_{n-1},y_n;s_{n-1}-s_n)f_n(y_n)
u_0(y_n,s_n)\,dy_1ds_1\dots dy_nds_n
\ea
for $n=1,2,\dots$, where $f_j\in L^q(\omega)$ ($j=1,\dots,n$). The Rytov series is expressed as
\be
\psi(x,t)=\sum_{n=1}^{\infty}(J_n\eta\otimes\cdots\otimes\eta)(x,t),
\ee
where
\ba
&
\left(J_n\eta\otimes\dots\otimes\eta\right)(x,t)=
\sum_{m=1}^n\frac{1}{mu_0(x,t)^n}
\\
&\times
\sum_{j_1+\cdots+j_m=n}\left((K_{j_1}\eta\otimes\dots\otimes\eta)(x,t)\right)
\times\cdots\times
\left((K_{j_m}\eta\otimes\dots\otimes\eta)(x,t)\right).
\ea

\begin{lem}
For $n=1,2,\dots$, $\|(1/u_0)K_n\|\le\nu\mu^{n-1}$.
\end{lem}

\begin{proof}
Using H\"{o}lder's inequality, we have
\ba
&
\left\|\frac{1}{u_0}K_nf_1\otimes\cdots\otimes f_n\right\|_{L^p(\pp\Omega)}^p
\\
&=
\left(\alpha_0^n\right)^p\int_{\pp\Omega}\Biggl|
\int_0^T\int_{\Omega}\cdots\int_0^T\int_{\Omega}
G(x,y_1;t-s_1)G(y_1,y_2;s_1-s_2)\cdots G(y_{n-1},y_n;s_{n-1}-s_n)
\\
&\times
\frac{u_0(y_n,t)}{u_0(x,t)}f_1(y_1)\cdots f_n(y_n)\,dy_1ds_1\cdots dy_nds_n\Biggr|^p\,dx
\\
&\le
\alpha_0^{np}\int_{\pp\Omega}\Biggl|\left(\int_{\Omega}\cdots\int_{\Omega}
|f_1(y_1)\cdots f_n(y_n)|^q\,dy_1\cdots dy_n\right)^{1/q}
\Biggl(\int_{\Omega}\cdots\int_{\Omega}
\\
&
\left|\int_0^T\cdots\int_0^TG(x,y_1;t-s_1)G(y_1,y_2;s_1-s_2)\cdots G(y_{n-1},y_n;s_{n-1}-s_n)\frac{u_0(y_n,t)}{u_0(x,t)}\,ds_1\dots ds_n\right|^{q/(q-1)}
\\
&dy_1\dots dy_n\Biggr)^{1-1/q}
\Biggr|^p\,dx
\\
&\le
\alpha_0^{np}\|f_1\|_{L^q(\Omega)}^p\cdots\|f_n\|_{L^q(\Omega)}^p
\int_{\pp\Omega}\left|\sup_{y_1,y_n\in\Omega\atop 0<t<T,\;0<s<T}G(x,y_1;s)\frac{u_0(y_n,t)}{u_0(x,t)}\right|^p\,dx
\\
&\times
\left(\int_{\Omega}\cdots\int_{\Omega}\left|\int_0^T\cdots\int_0^T
G(y_1,y_2;s_1-s_2)\cdots G(y_{n-1},y_n;s_{n-1}-s_n)\,ds_1\dots ds_n\right|^{q/(q-1)}
\,dy_1\cdots dy_n\right)^{p(1-1/q)}.
\ea
We define
\ba
\xi_{n-1}
&=
\alpha_0^{n-1}\Biggl(\int_{\Omega}\cdots\int_{\Omega}
\left|\int_0^T\cdots\int_0^TG(y_1,y_2;s_1-s_2)\cdots G(y_{n-1},y_n;s_{n-1}-s_n)\,ds_1\dots ds_n\right|^{q/(q-1)}
\\
&\,dy_1\cdots dy_n\Biggr)^{1-1/q}.
\ea
We have
\be
\xi_{n-1}\le\mu \xi_{n-2},\quad \xi_1\le T|\omega|^{1-1/q}\mu.
\ee
Hence,
\be
\xi_{n-1}\le\mu^{n-1}T|\omega|^{1-1/q}\quad(n=2,3,\dots).
\ee
We obtain
\be
\left\|\frac{1}{u_0}K_nf_1\otimes\cdots\otimes f_n\right\|_{L^p(\pp\Omega)}^p\le
\|f_1\|_{L^q(\omega)}^p\cdots\|f_n\|_{L^q(\omega)}^p\nu^p\mu^{p(n-1)}.
\ee
Therefore,
\be
\left\|\frac{1}{u_0}K_n\right\|
=
\sup_{f_1,\dots,f_n\in L^q(\omega)\atop f_i\neq0\;(i=1,\dots,n)}
\frac{\left\|\frac{1}{u_0}K_nf_1\otimes\cdots\otimes f_n\right\|_{L^p(\pp\Omega)}}
{\|f_1\|_{L^q(\omega)}\cdots\|f_n\|_{L^q(\omega)}}
\le\nu\mu^{n-1}.
\ee
\end{proof}

\begin{lem}
For $n=1,2,\dots$, $\|J_n\|\le\nu(\mu+\nu)^{n-1}$.
\label{estJn}
\end{lem}

\begin{proof}
\ba
\|J_n\|
&\le
\sum_{m=1}^n\frac{1}{m}\sum_{j_1+\cdots+j_m=n}
\left\|\frac{1}{u_0}K_{j_1}\right\|\cdots\left\|\frac{1}{u_0}K_{j_m}\right\|
\\
&\le
\sum_{m=1}^n\frac{1}{m}\sum_{j_1+\cdots+j_m=n}\nu^m\mu^{n-m}
\\
&\le
\sum_{m=1}^n\begin{pmatrix}n-1\\m-1\end{pmatrix}\nu^m\mu^{n-m}
\\
&=
\nu(\nu+\mu)^{n-1}.
\ea
\end{proof}

We obtain the inverse Rytov series by inverting the Rytov series:
\be
\eta=\mathcal{J}_1\psi+\mathcal{J}_2\psi\otimes\psi+\mathcal{J}_3\psi\otimes\psi\otimes\psi+\cdots.
\ee
Here, $\mathcal{J}_1$ is a regularized pseudoinverse of $J_1$ and for $n\ge2$,
\be
\mathcal{J}_n=-\left(\sum_{m=1}^{n-1}\mathcal{J}_m\sum_{j_1+\cdots+j_m=n}
J_{j_1}\otimes\cdots\otimes J_{j_m}\right)
\mathcal{J}_1\otimes\cdots\otimes\mathcal{J}_1.
\ee

\begin{lem}
We assume that there exists $M\in(0,1)$ such that $(\mu+2\nu)\|\mathcal{J}_1\|\le M$. Then $\|\mathcal{J}_n\|\le\exp(1/(1-M))(\mu+2\nu)^n\|\mathcal{J}_1\|$ for $n=2,3,\dots$.
\end{lem}

\begin{proof}
Using Lemma \ref{estJn}, we have
\ba
\|\mathcal{J}_n\|
&=
\left\|\left(\sum_{m=1}^{n-1}\mathcal{J}_m\sum_{j_1+\cdots+j_m=n}J_{j_1}\otimes\cdots\otimes J_{j_m}\right)\mathcal{J}_1\otimes\cdots\otimes\mathcal{J}_1\right\|
\\
&\le
\sum_{m=1}^{n-1}\|\mathcal{J}_m\|\begin{pmatrix}n-1\\m-1\end{pmatrix}
\nu^m(\mu+\nu)^{n-m}
\|\mathcal{J}_1\|^n
\\
&\le
\|\mathcal{J}_1\|^n\left(\sum_{m=1}^{n-1}\|\mathcal{J}_m\right)
\left(\sum_{m=1}^{n-1}\begin{pmatrix}n-1\\m-1\end{pmatrix}\nu^m(\mu+\nu)^{n-m}\right)
\\
&=
\|\mathcal{J}_1\|^n\left(\sum_{m=1}^{n-1}\|\mathcal{J}_m\|\right)
\left(\nu(\mu+2\nu)^{n-1}-\nu^n\right)
\\
&\le
\|\mathcal{J}_1\|^n(\mu+2\nu)^n\sum_{m=1}^{n-1}\|\mathcal{J}_m\|.
\ea
Thus,
\be
\|\mathcal{J}_n\|\le c_n\left((\mu+2\nu)\|\mathcal{J}_1\|\right)^n\|\mathcal{J}_1\|,
\ee
where $c_n$ are recursively given by
\be
c_{n+1}=c_n+\left((\mu+2\nu)\|\mathcal{J}_1\|\right)^nc_n,\quad c_2=1.
\ee
We obtain
\be
c_n=\prod_{m=2}^{n-1}\left(1+((\mu+2\nu)\|\mathcal{J}_1\|)^m\right),\quad n\ge3.
\ee
Hence for $n\ge3$,
\ba
\ln{c_n}
&=
\sum_{m=2}^{n-1}\ln\left(1+((\mu+2\nu)\|\mathcal{J}_1\|)^m\right)
\\
&\le
\sum_{m=2}^{n-1}\left((\mu+2\nu)\|\mathcal{J}_1\|\right)^m
\\
&\le
\frac{1}{1-(\mu+2\nu)\|\mathcal{J}_1\|}
\\
&\le
\frac{1}{1-M}.
\ea
We obtain $c_n\le e^{1/(1-M)}$ ($n\ge3$). We note that $1<e^{1/(1-M)}$. Thus the lemma is proved.
\end{proof}

\begin{thm}
We assume that there exists $M\in(0,1)$ such that $(\mu+2\nu)\|\mathcal{J}_1\|\le M$. Then the inverse Rytov series converges if $(\mu+2\nu)\|\psi\|_{L^p(\pp\Omega)}<1$.
\end{thm}

\begin{proof}
We have
\be
\|\eta\|_{L^q(\omega)}=
\sum_{n=1}^{\infty}\|\mathcal{J}_n\psi^{\otimes n}\|_{L^q(\omega)}\le
\sum_{n=1}^{\infty}\|\mathcal{J}_n\|\|\psi\|_{L^p(\pp\Omega)}^n\le
e^{\frac{1}{1-M}}\|\mathcal{J}_1\|\sum_{n=1}^{\infty}(\mu+2\nu)^n\|\psi\|_{L^p(\pp\Omega)}^n.
\ee
Thus the proof is complete.
\end{proof}

\section{Phantom experiments}
\label{phantomexp}

\begin{figure}[htbp]
\centering
\includegraphics[width=0.36\textwidth]{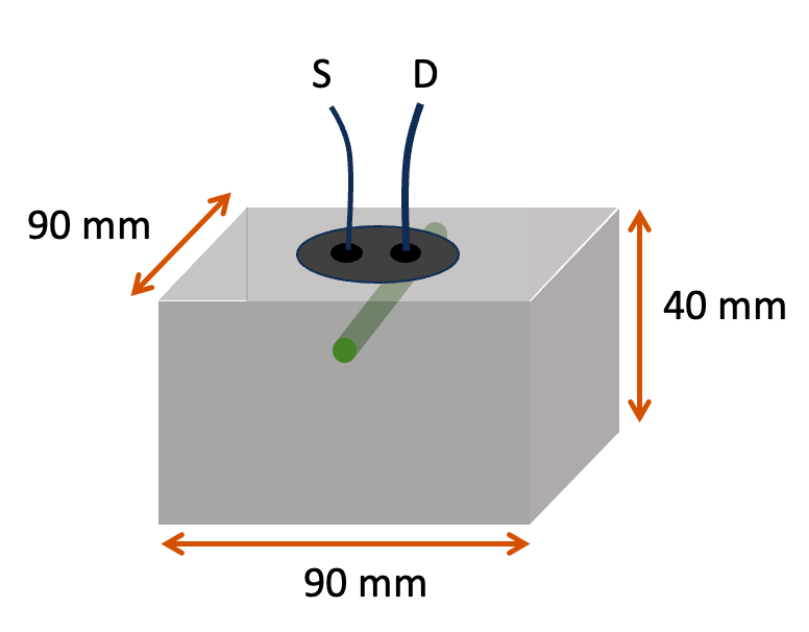}
\includegraphics[width=0.36\textwidth]{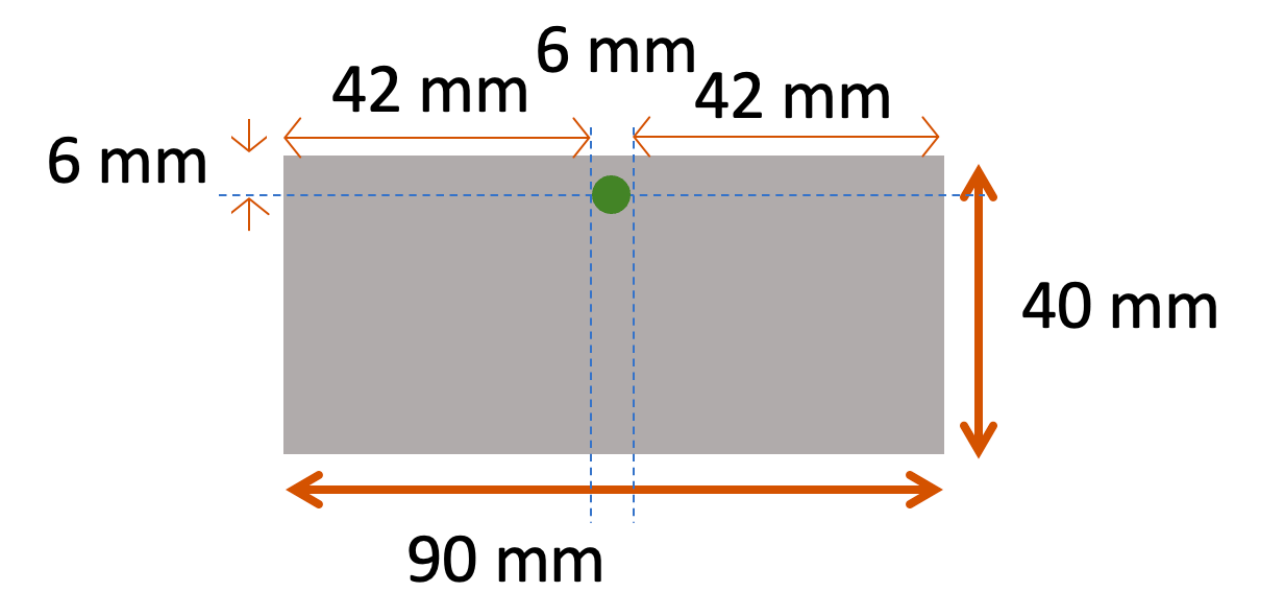}
\hspace{1mm}
\includegraphics[width=0.2\textwidth]{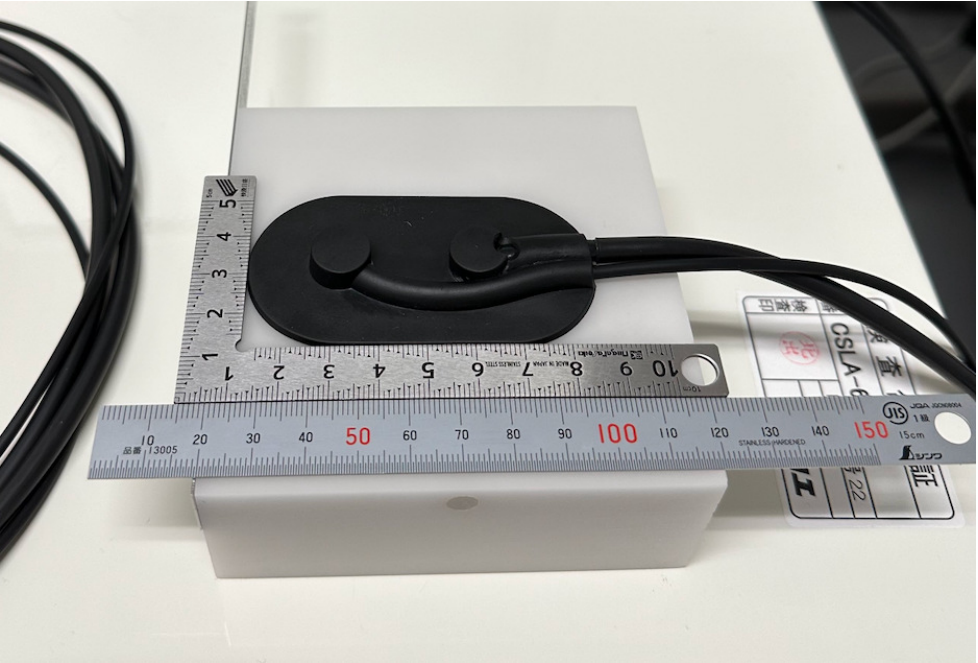}
\caption{\label{phantomfig}
The phantom. (Left) Schematic figure of the phantom experiment. (Middle) A cross section of the phantom. (Right) The experimental setup.}
\end{figure}

To test the proposed algorithm, the following phantom experiment was performed using TRS-21 (Hamamatsu Photonics K.~K., Japan). A solid phantom with an absorber rod (INO, Canada) was prepared and optical fibers for the source and detector were attached with a holder as shown in the left panel of Fig.~\ref{phantomfig}. As shown in the middle panel of Fig.~\ref{phantomfig}, the absorber rod of diameter $6\,{\rm mm}$ was embedded $6\,{\rm mm}$ below the surface of the phantom. The right panel of Fig.~\ref{phantomfig} shows a photo of the actual experiment setup.

The optical properties of the phantom were $\mu_s'=1\,{\rm mm}^{-1}$, $\mu_a=0.01\,{\rm mm}^{-1}$, and $\mathfrak{n}=1.51$ ($\mathfrak{n}$ is the refractive index of the phantom). The absorber rod shares the same $\mu_s'$ and $\mathfrak{n}$ but $\mu_a=0.03\,{\rm mm}^{-1}$ (This means $\eta=2$). 

Given that the absorber rod penetrated the phantom in one direction (from front to back), the holder was moved in another direction (from left to right, perpendicular to the rod) with a pitch $3\,{\rm mm}$. That is, time-resolved photon counts were taken three times at one position and the holder was moved $3\,{\rm mm}$ to the right. At each position, the average of three measurements was taken. There were nine measurement points ($M_{\rm SD}=9$): $x_{s1}^{(1)}=0$, $x_{s1}^{(2)}=3\,{\rm mm}$, $x_{s1}^{(3)}=6\,{\rm mm}$ , $x_{s1}^{(4)}=9\,{\rm mm}$, $x_{s1}^{(5)}=12\,{\rm mm}$, $x_{s1}^{(6)}=15\,{\rm mm}$, $x_{s1}^{(7)}=18\,{\rm mm}$, $x_{s1}^{(8)}=21\,{\rm mm}$, and $x_{s1}^{(9)}=24\,{\rm mm}$ both for the SD distance $x_{d1}^{(i)}-x_{s1}^{(i)}=2\,{\rm cm}$ and $3\,{\rm cm}$ ($i=1,\dots,M_{\rm SD}$).

At each point $(x_{s1}^{(i)},x_{d1}^{(i)})$ ($i=1,\dots,M_{\rm SD}$), arrived photons are scored every $10\,{\rm ps}$ with $T=10.23\,{\rm ns}$. To obtain $\Gamma_0,\Gamma$, we took the moving average for every temporal profile by taking five points before and after each time (average of 11 temporal points). Here, $\Gamma_0$ is the reflectance which corresponds to $u_0$. 

In each temporal profile, we considered the time period during $t_1\le t\le t_2$ for which the peak of the temporal profile lies. We set $t_1=0.25\,{\rm ns}$, $t_2=0.45\,{\rm ns}$. Indeed, the peak of the source $h(t)$ is approximately at $t_1$. Define
\be
\Psi(x_d^{(i)},x_s^{(i)};t)=
\ln\frac{\Gamma_0^{(i)}(t)}{\Gamma^{(i)}(t)},\quad t_1\le t\le t_2,
\ee
where $\Gamma_0^{(i)}(t),\Gamma^{(i)}(t)$ are reflectances $\Gamma_0,\Gamma$ for the $i$th SD (source-detector) pair. With unknown positive constants $C_{0i},C_i$, we can write
\ba
\Gamma_0^{(i)}(t)&=C_{0i}w_0(x_d^{(i)},t;x_s^{(i)}),
\\
\Gamma^{(i)}(t)&=C_iw(x_d^{(i)},t;x_s^{(i)})
\ea
for $t_1\le t\le t_2$. The constants $C_{0i},C_i$ depend on experimental conditions and in many cases take different values for different $i$. We have
\be
\Psi(x_d^{(i)},x_s^{(i)};t)=
\ln\frac{w_0(x_d^{(i)},t;x_s^{(i)})}{w(x_d^{(i)},t;x_s^{(i)})}+\ln\frac{C_{0i}}{C_i}=
\psi(x_d^{(i)},x_s^{(i)};t)+\ln\frac{C_{0i}}{C_i}.
\ee
Let us define
\ba
\phi(x_d^{(i)},x_s^{(i)};t)
&=
\Psi(x_d^{(i)},x_s^{(i)};t+\tau)-\Psi(x_d^{(i)},x_s^{(i)};t)
\\
&=
\psi(x_d^{(i)},x_s^{(i)};t+\tau)-\psi(x_d^{(i)},x_s^{(i)};t),
\ea
where $\tau>0$ is a constant. We set $\tau=20\,{\rm ps}$. Then we have
\be
\phi(x_d^{(i)},x_s^{(i)};t)=
I_1(\eta)(x_d^{(i)},x_s^{(i)};t)+I_2(\eta)(x_d^{(i)},x_s^{(i)};t)+\cdots,
\ee
where
\be
I_n(\eta)(x_d^{(i)},x_s^{(i)};t)=
J_n(\eta)(x_d^{(i)},x_s^{(i)};t+\tau)-J_n(\eta)(x_d^{(i)},x_s^{(i)};t)
\ee
for $n=1,2,\dots$.

We will reconstruct $\eta$ from the data $\phi(x_s^{(i)},t)$ using the series for $I_n(\eta)$.

\section{Reconstructed images}
\label{results}

\subsection{Time-resolved measurements}

\begin{figure*}[htbp]
\centering
\includegraphics[width=0.8\textwidth]{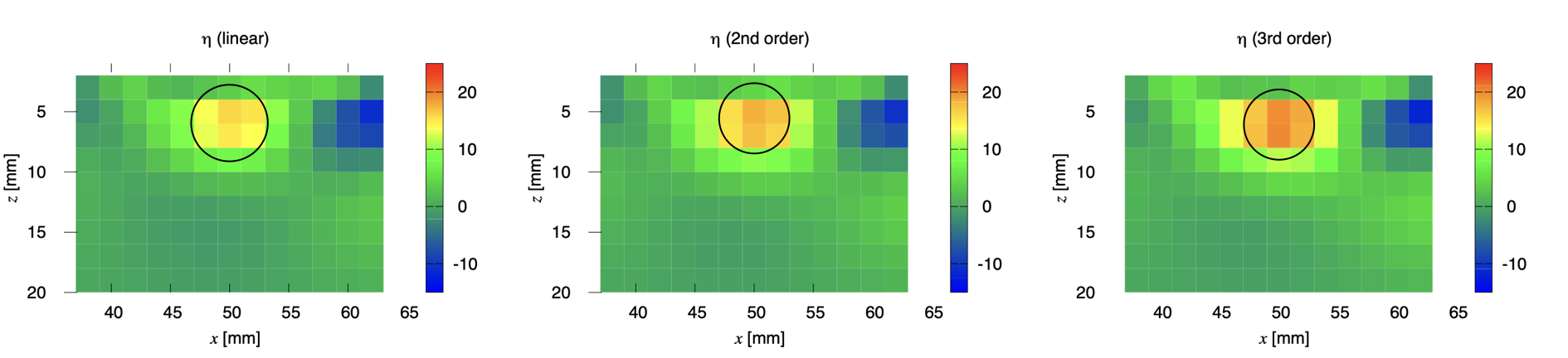}
\caption{\label{fig2a}
In the case of $d_{\rm SD}=2\,{\rm cm}$ and $\Delta x=2\,{\rm mm}$. Reconstructed images, from the left, $N=1$ (the conventional Rytov approximation), $N=2$, and $N=3$. The true position of the absorber rod is shown by a black circle.}
\end{figure*}

\begin{figure*}[htbp]
\centering
\includegraphics[width=0.8\textwidth]{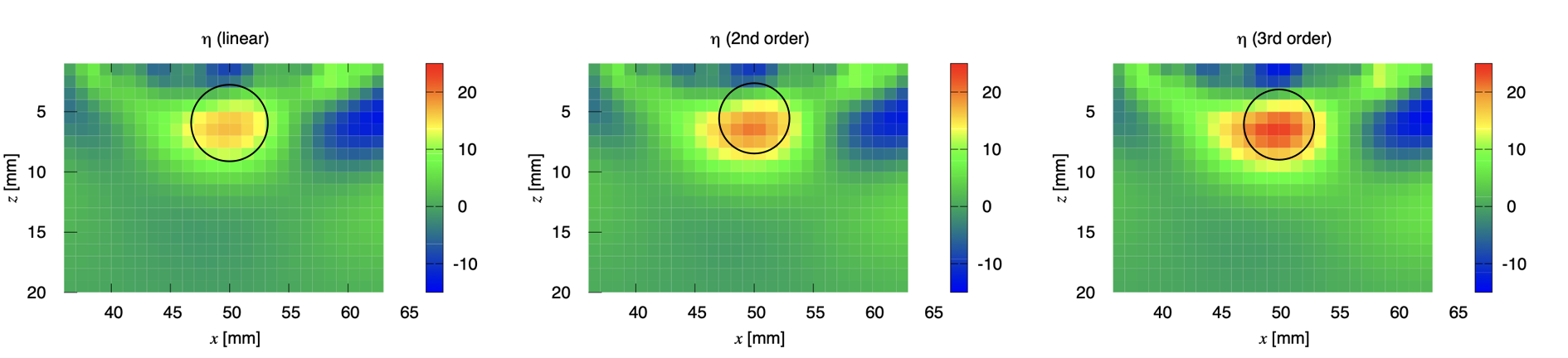}
\caption{\label{fig2b}
In the case of $d_{\rm SD}=2\,{\rm cm}$ and $\Delta x=1\,{\rm mm}$. Reconstructed images, from the left, $N=1$ (the conventional Rytov approximation), $N=2$, and $N=3$. The true position of the absorber rod is shown by a black circle.}
\end{figure*}

\begin{figure*}[htbp]
\centering
\includegraphics[width=0.8\textwidth]{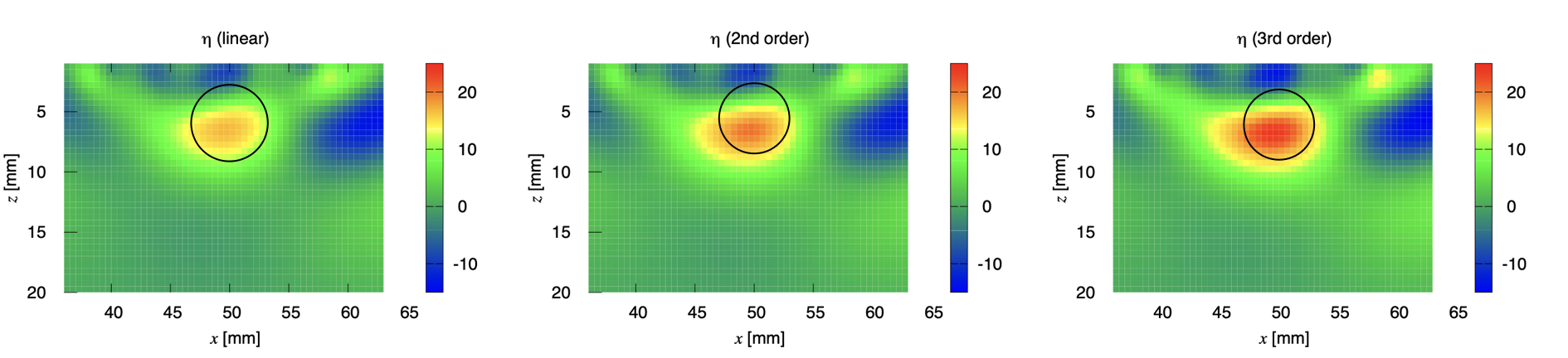}
\caption{\label{fig2c}
In the case of $d_{\rm SD}=2\,{\rm cm}$ and $\Delta x=0.5\,{\rm mm}$. Reconstructed images, from the left, $N=1$ (the conventional Rytov approximation), $N=2$, and $N=3$. The true position of the absorber rod is shown by a black circle.}
\end{figure*}

\begin{figure*}[htbp]
\centering
\includegraphics[width=0.8\textwidth]{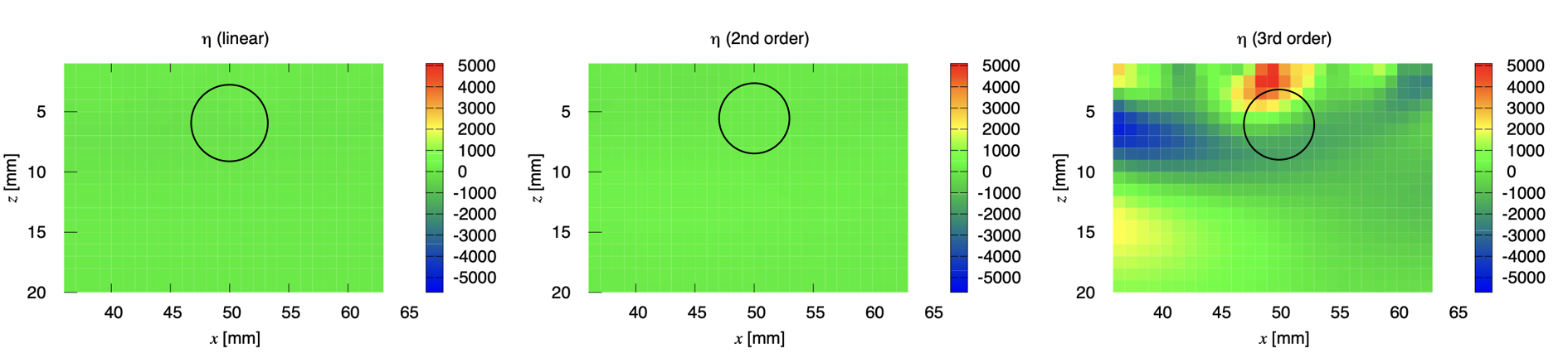}
\caption{\label{fig3}
In the case of $d_{\rm SD}=3\,{\rm cm}$ and $\Delta x=1\,{\rm mm}$. Reconstructed images, from the left, $N=1$ (the conventional Rytov approximation), $N=2$, and $N=3$. The true position of the absorber rod is shown by a black circle.}
\end{figure*}

Reconstructed images are shown in Figs.~\ref{fig2a} through \ref{fig3}. In each figure, the left panel shows the linear reconstruction ($N=1$), i.e., the conventional Rytov approximation, the center panel shows the reconstruction for $N=2$, and the right panel is the reconstructed image when $N=3$. The black circle in each panel shows the true position of the absorber rod. In all cases, $9$ largest singular values were used. The SD distance $d_{\rm SD}$ was $2\,{\rm cm}$ for Figs.~\ref{fig2a}, \ref{fig2b}, and \ref{fig2c}. In Fig.~\ref{fig3}, $d_{\rm SD}=3\,{\rm cm}$. For the resolution, $\Delta x=2\,{\rm mm}$ in Fig.~\ref{fig2a}, $\Delta x=1\,{\rm mm}$ in Fig.~\ref{fig2b}, $\Delta x=0.5\,{\rm mm}$ in Fig.~\ref{fig2c}, and $\Delta x=1\,{\rm mm}$ for Fig.~\ref{fig3}.

\subsection{Dependence on the target size}
\label{size}

\begin{figure*}[htbp]
\centering
\includegraphics[width=0.3\textwidth]{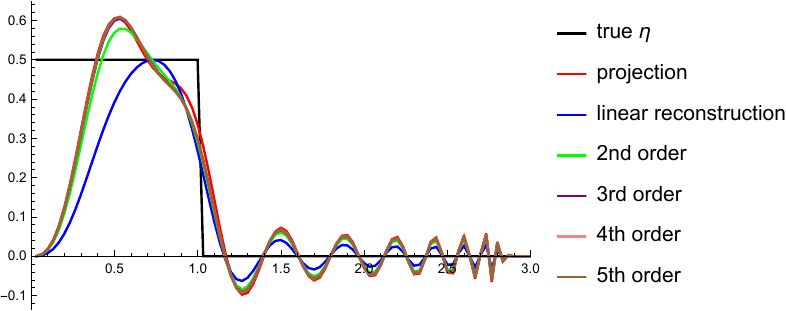}
\includegraphics[width=0.3\textwidth]{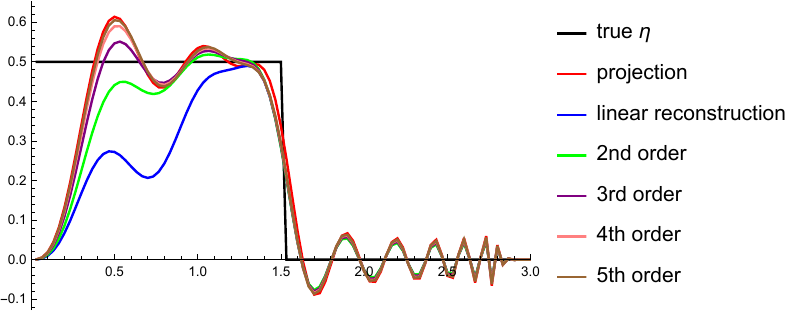}
\includegraphics[width=0.3\textwidth]{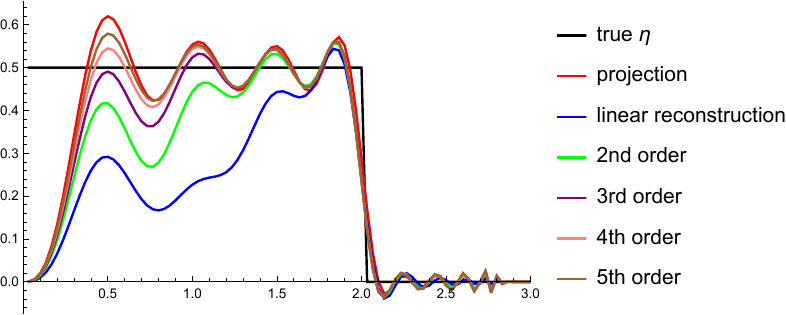}
\caption{\label{fig4}
The radius $r_a$ of the target is, from the left, $1$, $1.5$, and $2$. The parameter $\eta_a$ is set to $0.5$.
}
\end{figure*}

Here we consider the radial problem in Sec.~\ref{tindep}. Let us set $D_0=1$, $\alpha_0=1$, $R=3$, $\ell=0.3$ $\eta_a=0.5$. We take $r_a=1$, $1.5$, $2$. For these cases, we set $\sigma_0$ such that 23 singular values are used for reconstruction. Reconstructed figures are shown in Fig.~\ref{fig4}. Since the inverse problem is ill-posed, the projection $\mathcal{J}_1J_1\eta$ is the best reconstruction within the regularization. In all panels of Fig.~\ref{fig4}, the ground truth $\eta$ and projection are shown in black and red, respectively. The linear reconstruction (the conventional Rytov approximation) $\bv{\eta}^{(1)}$ is shown in blue. Moreover, the light green, purple, coral, and ocher lines denote $\bv{\eta}^{(2)}$, $\bv{\eta}^{(3)}$, $\bv{\eta}^{(4)}$, and $\bv{\eta}^{(5)}$, respectively. In Fig.~\ref{fig4}, when $r_a=1$ and the support of $\eta$ is small (the left panel), the third order reconstruction $\bv{\eta}^{(3)}$ already gives an almost converged value. When $r_a=2$ and the support of $\eta$ is large (the right panel), the fifth order reconstruction $\bv{\eta}^{(5)}$ is close but still different from the projection. In the intermediate case when $r_a=1.5$ (the center panel), the fifth order reconstruction $\bv{\eta}^{(5)}$ is almost identical to the projection.

\section{Concluding remarks}
\label{concl}

Since the ratio of the absorption coefficients inside and outside the absorber rod is large ($\mu_a=0.03\,{\rm mm}^{-1}$ and $0.01\,{\rm mm}^{-1}$) and $\eta=2$, the inverse series does not converge and the reconstruction of the value of $\eta$ is difficult. However, Figs.~\ref{fig2a}, \ref{fig2b}, and \ref{fig2c} show that the target is more clearly reconstructed if nonlinear terms are added.

We found in \cite{Machida-Osada-Kagawa} that the depth of the center of the banana is $d_{\rm SD}/(2\sqrt{2})$, where $d_{\rm SD}$ is the distance between the source and detector. According to this formula, the depth of the banana is $7.1\,{\rm mm}$. When the SD distance $3\,{\rm cm}$, the depth of the banana is $10.6\,{\rm mm}$, which is deeper than the depth of the absorber rod. This explains why the reconstruction in Fig.~\ref{fig3} was not successful.

When the absorber rod was reconstructed for the phantom experiment, it was assumed a priori that the rod penetrated the phantom. Three dimensional reconstruction is necessary unless this fact is used. In this case, the formulation developed in this paper can be extended in a straightforward manner.

We note that in many cases diffuse optical tomography in time domain can be formulated using the time-independent diffusion equation by the Laplace or Fourier transform. In this case, the reconstruction can be done using the inverse Rytov series for the time-independent diffusion equation as was developed in Sec.~\ref{tindep}.

In \cite{Machida23}, the inverse Rytov series was first constructed for the time-independent diffusion equation. In this paper, we have also developed the inverse Rytov series in time domain. To handle experimental time-resolved data, we considered the forward series by the subtraction of the Rytov series. Then the corresponding inverse series could be computed recursively. We used nonlinear Rytov approximations which are proposed in this paper to obtain reconstructed images for the phantom experiment. By this, the use of the inverse Rytov series was demonstrated. In addition, the dependence of the reconstruction on $|\omega|$ was numerically studied in Secs.~\ref{tindep} and \ref{size}.

\vskip\baselineskip
\noindent {\bf Acknowledgments.}
This work was supported by JST, PRESTO Grant Number JPMJPR2027.



\end{document}